\documentstyle[aps]{revtex}
\begin{document}
\voffset 0.8truecm
\title{Remarks on entanglement assisted classical capacity
}
\author{
Heng Fan
}
\address{
Quantum computation and information project,
ERATO, 
Japan Science and Technology Corporation,\\
Daini Hongo White Building 201, Hongo 5-28-3, Bunkyo-ku, Tokyo 133-0033, Japan.\\
}
\maketitle

\begin{abstract}
The property of the optimal signal ensembles of entanglement
assisted channel capacity is studied.
A relationship between entanglement assisted channel
capacity and one-shot capacity of unassisted channel
is obtained.
The data processing inequalities, convexity and additivity of the
entanglement assisted channel capacity are reformulated by
simple methods.
\end{abstract}
       
\pacs{03.67.Lx, 03.65.Ta, 32.80.Qk}

Capacities of quantum channels are the basic problems in
quantum information theory. 
The entanglement assisted classical capacity $C_E$ is the maximum
asympototic rate of reliable bit transmission with the
help of unlimited prior entanglement between the sender (Alice)
and receiver (Bob). Bennett, Shor, Smolin and Thapliyal (BSST) recently
proved the entanglement assisted classical capacity is the
maximual quantum mutual information between Alice and Bob, which
we call it BSST theorem in this paper\cite{BSST,BSST1},
\begin{eqnarray}
C_E({\cal {N}})=\max _{\rho \in {\cal {H}}_{in}}
S(\rho _A)+S({\cal {N}}(\rho _{A}))
-S({\cal {N}}\otimes I(|\Phi _{AB}\rangle
\langle \Phi _{AB}|)),
\label{BSST}
\end{eqnarray}
where $S(\rho )=-{\rm Tr}\rho \log \rho $ is the
von Neumann entropy,
$|\Phi _{AB}\rangle $ is the purification of input state
$\rho _A $. Note that all purifications of $\rho _A$ give the
same entropy in this formula.
Holevo subsequently gave
a simple proof\cite{H} of BSST theorem.
Before BSST's work, Adami and Cerf already
characterized some important properties of quantum mutual
information\cite{AC}. In this paper, we revisit
some of these properties in the framework
of entanglement assisted classical capacity.
In BSST theorem, we need the
purifiction state $|\Phi _{AB}\rangle $ of the input
state $\rho _A$ which is not necessarily a pure state in
the formula of quantum mutual information.
We give a simple proof of the data processing inequalities,
and convexity property for
the entanglement assisted channel capacity. We also characterize
one property of the optimal signal ensembles. A simple proof
is given for the additivity of entanglement assisted
channel capacity.
A relationship between entanglement assisted channel
capacity and one-shot unassisted capacity is obtained which is
more exact than previous result.

The data processing inequalities are already known for
classical information theory\cite{CT}, and for the
Holevo-Schumacher-Westmoreland classical capacity of
the quantum channel\cite{H0,SW,NC}. 
We present the quantum data processing inequalities
for entanglement assisted channel capacity,
\begin{eqnarray}
C_E({\cal {N}}_2\circ {\cal {N}}_1)
\le \min \{ C_E({\cal {N}}_1),
C_E({\cal {N}}_2)\} ,
\label{datap}
\end{eqnarray}
where ${\cal {N}}_1$ and ${\cal {N}}_2$ are two discrete
memoryless quantum channels such that the input of 
${\cal {N}}_2$ is the output of ${\cal {N}}_1$.

The entanglement assisted classical capacity can be formulated
by quantum relative entropy
\begin{eqnarray}
C_E({\cal {N}})=\max _{\rho \in {\cal {H}}_{in}}
S({\cal {N}}\otimes I(|\Phi _{AB}\rangle
\langle \Phi _{AB}|)||{\cal {N}}(\rho _{A})\otimes \rho _B),
\label{relative}
\end{eqnarray}
where the relative entropy is defined as
$S\left( \rho ||\sigma \right)
\equiv {\rm Tr}\rho ({\rm log}\rho -{\rm log}\sigma )$, $\rho $
and $\sigma $ are density operators.
In formula (\ref{relative}), we also used the fact that for pure
state $|\Phi _{AB}\rangle $, $S(\rho _A)=S(\rho _B)$.
We know the quantum relative entropy has the property
of monotonicity under completely positive, trace perserving (CPTP)
maps\cite{NC,Ruskai}.
Since ${\cal {N}}_2$ is the CPTP map, so does
${\cal {N}}_2 \otimes I$, we know
\begin{eqnarray}
S({\cal {N}}_1\otimes I(|\Phi _{AB}\rangle
\langle \Phi _{AB}|)||{\cal {N}}_1(\rho _{A})\otimes \rho _B)
\ge 
S({\cal {N}}_2\circ {\cal {N}}_1\otimes I(|\Phi _{AB}\rangle
\langle \Phi _{AB}|)||{\cal {N}}_2\circ {\cal {N}}_1
(\rho _{A})\otimes \rho _B).
\end{eqnarray}
Thus we know
\begin{eqnarray}
C_E({\cal {N}}_1)\ge
C_E({\cal {N}}_2\circ {\cal {N}}_1).
\label{onehalf}
\end{eqnarray}
This is a much simpler proof. So, we proved half of the data
processing inequalities (\ref{datap}).

Next, let us prove the second half of the relation (\ref{datap}).
What we need to prove is the following inequality,
\begin{eqnarray}
&&S\left( \rho _A\right) +S\left( {\cal {N}}_2\circ {\cal {N}}_1(\rho _A)
\right) -S\left( {\cal {N}}_2\circ {\cal {N}}_1\otimes
I(|\Phi _{AB}\rangle \langle \Phi _{AB}|)\right)
\nonumber \\
&\le &S\left( {\cal {N}}_1(\rho _A)\right)
+
S\left( {\cal {N}}_2\circ {\cal {N}}_1(\rho _A)
\right) -S\left( {\cal {N}}_2\otimes
I(|\Phi _{A'B'}\rangle \langle \Phi _{A'B'}|)\right) ,
\label{reverse}
\end{eqnarray}
where $|\Phi _{A'B'}\rangle $ is a purification of
${\cal {N}}_1(\rho _A)\equiv \rho _{A'}$,
the output
state of channel ${\cal {N}}_1$.
A general quantum channel can be realized by adding first
an ancilla state, performing next the unitary transformation
corresponding to the quantum channel and finally
tracing out the ancilla system.
We suppose the channel ${\cal {N}}_1$ can be realized by
unitary transformation $U_{AC}$, where $C$ is the ancilla system.
The whole system after unitary transformation can be written as
$|\Phi _{A'C'B}\rangle
=U_{AC}\otimes I_B|\Phi _{AB}\rangle |0\rangle _C$,
and we know
$\rho _{A'}={\cal {N}}_1(\rho _A)
={\rm Tr}_{C'B}|\Phi _{A'C'B}\rangle \langle
\Phi _{A'C'B}|$.
Since $|\Phi _{A'C'B}\rangle $ is already a purification of
${\cal {N}}_1(\rho _A)$, we simply define
$|\Phi _{A'B'}\rangle \equiv |\Phi _{A'C'B}\rangle $,
and identify $C'B$ as the reference state $B'$ in the
purification.  So, the inequality (\ref{reverse}) becomes
the following,
\begin{eqnarray}
S(\rho _B)-S(\rho _{A''B})
\le S(\rho _{C'B})-S(\rho _{A''C'B}) ,
\end{eqnarray}
where we denote $\rho _{A''C'B}={\cal {N}}_2\otimes I_{C'B}
(|\Phi _{A'C'B}\rangle \langle \Phi _{A'C'B})$.
This inequality is the strong subadditivity inequality
\cite{Ruskai}. So we proved the relation (\ref{reverse}).
Considering the definition of the entanglement assisted channel capacity
(\ref{BSST}), we know
\begin{eqnarray}
C_E({\cal {N}}_2\circ {\cal {N}}_1)
\le C_E({\cal {N}}_2).
\label{twohalf}
\end{eqnarray}
Combining (\ref{onehalf}) and
(\ref{twohalf}), we proved the data processing inequalities
(\ref{datap}).
We remark that for entanglement
assisted channel capacity, the output of the
the optimal input $\rho _A$ for channel ${\cal {N}}_1$ is not
necessarily the optimal input for channel ${\cal {N}}_2$,
similary the optimal input for channel ${\cal {N}}_1$ is
not necessarily the optimal input for channel
${\cal {N}}_2\circ {\cal {N}}_1$. 

The convexity of the entanglement assisted channel capacity
is written as,
\begin{eqnarray}
C_E(\sum _ip_i{\cal {N}}_i)
\le \sum _ip_iC_E({\cal {N}}_i).
\label{convex}
\end{eqnarray}
It means the capacity of a channel which is an average of
several channels cannot exceed the average capacity
of these channels.
Since the entanglement assisted channel capacity can be
formulated by relative entropy (\ref{relative}), with the help of
the joint convexity of quantum relative entropy, we can
obtain (\ref{convex}) straightforwardly,
\begin{eqnarray}
C_E(\sum _ip_i{\cal {N}}_i)
&=&\max _{\rho \in {\cal {H}}_{in}}
S(\sum _ip_i{\cal {N}}_i\otimes I(|\Phi _{AB}\rangle
\langle \Phi _{AB}|)||\sum _ip_i{\cal {N}}_i(\rho _{A})\otimes \rho _B)
\nonumber \\
&\le &\max _{\rho \in {\cal {H}}_{in}}
\sum _pp_iS({\cal {N}}_i\otimes I(|\Phi _{AB}\rangle
\langle \Phi _{AB}|)||{\cal {N}}_i(\rho _{A})\otimes \rho _B).
\end{eqnarray}
So, we give a simple proof of the data processing inequalities,
and convexity of entanglement assisted channel capacity,
some results appeared essentially in Ref.\cite{SN,BNS,AC}. 

Some important properties characterizing the 
the optimal signal ensembles of Holevo-Schumacher-Westmoreland channel
capacity were studied by Schumacher and Westmoreland\cite{SW1}.
In this paper, we would like to study some propertis of
optimal signal ensembles for entanglement assisted channel
capacity.

Suppose we have different optimal input density operators
$\rho _A^i\in {\cal {H}}_{in}$ for quantum channel ${\cal {N}}$,
that means the channel capacity $C_E({\cal {N}})$ can be
achieved by every input density operator $\rho _A^i$,
then the average density operator of $\rho _A^i$ with
arbitrary probability distribution $p_i$, $\sum _ip_i=1$,
is also an optimal input state.
Explicitly, we suppose
\begin{eqnarray}
C_E({\cal {N}})=S(\rho _A^i)+S({\cal {N}}(\rho _{A}^i))
-S({\cal {N}}\otimes I(|\Phi _{AB}^i\rangle
\langle \Phi _{AB}^i|)),
\label{suppose}
\end{eqnarray}
where $|\Phi _{AB}^i\rangle $ is the purification of
$\rho _A^i$. Then we should have the following result
\begin{eqnarray}
C_E({\cal {N}})=S(\sum _ip_i\rho _A^i)+S({\cal {N}}(\sum _ip_i\rho _{A}^i))
-S({\cal {N}}\otimes I(|\Phi _{AB}\rangle
\langle \Phi _{AB}|)),
\label{optimal}
\end{eqnarray}
where $|\Phi _{AB}\rangle $ is the purification of
$\rho_A=\sum _ip_i\rho _A^i$, i.e., $\rho _A$ is the optimal input
state.

{\it Proof}:
Since we have the input states $\rho _A^i\in {\cal {H}}_{in}$,
we know the average density operator satisfies
$\sum _ip_i\rho _A^i=\rho _A\in {\cal {H}}_{in}$.
According to the definition of entanglement assisted channel
capacity, we know the inequality $\ge $ holds
\begin{eqnarray}
C_E({\cal {N}})\ge
S(\sum _ip_i\rho _A^i)+S({\cal {N}}(\sum _ip_i\rho _{A}^i))
-S({\cal {N}}\otimes I(|\Phi _{AB}\rangle
\langle \Phi _{AB}|)).
\label{know}
\end{eqnarray}
So, in order to prove relation (\ref{optimal}), we just need to
prove the opposite inequality $\le $.
Because $|\Phi _{AB}^i\rangle $ is the purification of
$\rho _{A}^i$, the average density operator
$\rho _A=\sum _ip_i\rho _A^i$ has purification
$|\Phi _{ABD}\rangle \equiv \sum_i\sqrt{p_i}|\Phi ^i_{AB}\rangle
|i\rangle _D$,
where $BD$ is the reference system and
$\rho _A={\rm Tr}_{BD}(|\Phi _{ABD}\rangle \langle \Phi _{ABD}|)$ .
As previously, we add the ancilla system $C$, and use the
unitary transformation $U_{AC}$ to realize quantum channel
${\cal {N}}$. The relation is written as
\begin{eqnarray}
|\Phi _{A'C'BD}\rangle 
=\sum _i\sqrt{p_i}|\Phi _{A'C'B}^i\rangle |i\rangle _D .
\label{final}
\end{eqnarray}
And also we have the inequality, 
\begin{eqnarray}
S(\sum _ip_i\rho ^i_{A'})
+S(\sum _ip_i\rho ^i_{A'C'})-S(\sum _ip_i\rho _{C'})
\ge
\sum _ip_i
[S(\rho _{A'}^i)+S(\rho ^i_{A'C'})-S(\rho ^i_{C'})] .
\end{eqnarray}
But we know
\begin{eqnarray} 
S(\rho _{A'}^i)+S(\rho ^i_{A'C'})-S(\rho ^i_{C'})=
S\left({\cal {N}}(\rho _{A}^i)\right)
+S(\rho _A^i)-S\left( {\cal {N}}\otimes I_B(
|\Phi ^i_{AB}\rangle \langle \Phi _{AB}^i|)\right)
=C_E({\cal {N}}),
\end{eqnarray}
where the last equality is due to the fact that $\rho _A^i$
is the optimal input.
So, we prove the inequality
\begin{eqnarray}
S(\rho _A)+S\left( {\cal {N}}(\rho _A)\right)
-S\left( {\cal {N}}\otimes I_{BD}
(|\Phi _{ABD}\rangle \langle \Phi _{ABD}|)\right)
\ge \sum _ip_iC_E({\cal {N}})=C_E({\cal {N}}).
\end{eqnarray}
Thus we know relation (\ref{optimal}) holds.
Eqs.(\ref{suppose}) and (\ref{optimal}) are a consequence
of the concavity of the entanglement assisted channel capacity.
We remark that 
the concavity of quantum mutual information is proved in Ref.\cite{AC}.

The additivity problem is one of the most basic problems
in classical and quantum information theories.
The additivity of entanglement measures such as the widely
accepted entanglement of formation for mixed state
was only proved for a few cases and remains as a conjecture
for the general case. The additivity of
Holevo-Schumacher-Westmoreland channel
capacity is also a long-standing conjecture.
The situation for the additivity
of entanglement assisted channel capacity is different.
The additivity was essentially proved by Adami and Cerf\cite{AC}
before the appearance of the definition of entanglement assisted
channel capacity. Next, we would like to reformulate
this result with the help of joint subadditivity of the
conditional entropy proposed independently by Nielsen\cite{NC}
and Adami and Cerf\cite{AC}. We follow the formulae by Nielsen.

First let us derive the relation of joint subadditivity of the
conditional entropy. From the strong subadditivity of
quantum entropy, we know,
\begin{eqnarray}
S(\rho _{ABCD})+S(\rho _C)\le S(\rho _{AC})
+S(\rho _{BCD}).
\end{eqnarray}
Add $S(\rho _D)$ on both sides of the inequality, we have
\begin{eqnarray}
S(\rho _{ABCD})+S(\rho _C)+S(\rho _D)&\le &S(\rho _{AC})
+S(\rho _{BCD})+S(\rho _D)
\nonumber \\
&\le &
S(\rho _{AC})+S(\rho _{BD})+S(\rho _{CD}),
\label{joint}
\end{eqnarray}
where the last inequality is due to the strong subadditivity.

The additivity of entanglement assisted channel capacity means
the following,
\begin{eqnarray}
C_E({\cal {N}}_1\otimes {\cal {N}}_2)
=C_E({\cal {N}}_1)+C_E({\cal {N}}_2).
\label{add}
\end{eqnarray}
From the BSST theorem, we know straightforwardly that
inequality $\ge $ holds,
let's show the opposite inequality first proved by Adami and Cerf\cite{AC},
\begin{eqnarray}
&&S(\rho _{A_1A_2})-S\left(
{\cal {N}}_1\otimes {\cal {N}}_2\otimes I_B
(|\Phi _{A_1A_2B}\rangle \langle \Phi _{A_1A_2B}|)\right)
\nonumber \\
&\le & 
S(\rho _{A_1})+S(\rho _{A_2})-S\left(
{\cal {N}}_1\otimes I_{A_2B}
(|\Phi _{A_1A_2B}\rangle \langle \Phi _{A_1A_2B}|)\right)
-S\left(
{\cal {N}}_2\otimes I_{A_1B}
(|\Phi _{A_2A_1B}\rangle \langle \Phi _{A_2A_1B}|)\right) .
\end{eqnarray}
We can use ancilla system $C_1,C_2$ and unitary transformations
$U_{A_1C_1}$ and $U_{A_2C_2}$ to realize the quantum channels
${\cal {N}}_1$ and ${\cal {N}}_2$, and it can be written as
\begin{eqnarray}
|\Phi _{A_1'C_1A_2'C_2'B}\rangle
=(U_{A_1C_1}\otimes U_{A_2C_2}\otimes I_B)
|\Phi _{A_1A_2B}\rangle |0\rangle _{C_1}|0\rangle _{C_2}.
\end{eqnarray}
So, we should prove the following inequality
\begin{eqnarray}
S(\rho _{A_1'C_1'A_2'C_2'})-S\left(\rho _{C_1'C_2'}\right)
\le
S(\rho _{A_1'C_1'})+S(\rho _{A_2'C_2'})-S\left(\rho _{C_1'}\right)
-S\left(\rho _{C_2'}\right) .
\end{eqnarray}
This inequality is exactly the
joint subadditivity of the conditional entropy (\ref{joint}).
From the definition of entanglement assisted channel capacity,
and considering the subadditivity inequality
for $S( {\cal {N}}_1\otimes {\cal {N}}_2(\rho _{A_1A_2}))$,
we know the following inequality holds,
\begin{eqnarray}
C_E({\cal {N}}_1\otimes {\cal {N}}_2)
\le C_E({\cal {N}}_1)+C_E({\cal {N}}_2).
\end{eqnarray}
Thus, the entanglement assisted channel capacity is additive.

Holevo pointed out that the entanglement assisted channel
capacity is upper bounded by $\log d$ plus the unassisted
capacity\cite{H}, where $d$ is the dimension
of the input quantum state space.
It was recently proved that this upper bound
can be replaced as $\log d$ plus one-shot unassisted
channel capacity\cite{Fan}. This result is based mainly on the
fact that if $\rho _A$ has the pure states decomposition as
$\rho _A=\sum _ip_i|\Phi _A^i\rangle \langle \Phi _A^i|$,
then the following inequality holds,
\begin{eqnarray}
S({\cal {N}}\otimes I(|\Phi _{AB}\rangle
\langle \Phi _{AB}|))\ge \sum _ip_iS\left(
{\cal {N}}(|\Phi _A^i\rangle \langle \Phi _A^i|)\right) .
\end{eqnarray}
This inequality can be proved by several methods\cite{S,Fan,H1}.
Here, we will present a more exact result
with the help of the result obtained by Schumacher and
Westmoreland and Holevo \cite{SW2,H1}.

Suppose the quantum channel ${\cal {N}}$ can be
realized by the unitary transformation $U_{AC}$ with ancilla system
$C$. A purification of $\rho _A$ is written as
$|\Phi _{AB}\rangle =\sum _i\sqrt{p_i}|\Phi _A^i\rangle |i\rangle _B$.
Define $|\Phi _{A'C'B}\rangle =\sum _i\sqrt{p_i}
|\Phi ^i_{A'C'}\rangle |i\rangle _B$,
the entanglement assisted capacity of channel ${\cal {N}}$ is written as
\begin{eqnarray}
C_E({\cal {N}})
=\max _{\rho _A\in {\cal {H}}_{in}}
S(\sum _ip_i
|\Phi _A^i\rangle \langle \Phi _A^i|) +
S(\sum _ip_i{\cal {N}}(
|\Phi _A^i\rangle \langle \Phi _A^i|))
-S(\sum _ip_i\rho _{C'}^i).
\label{relation}
\end{eqnarray}
For pure state $|\Phi _{A'C'}^i\rangle $, we know
$S(\rho ^i_{A'})=S(\rho ^i_{C'})$.
Inserting the term $\sum _ip_i
[S(\rho ^i_{A'})-S(\rho ^i_{C'})]$ which is zero into the relation
(\ref{relation}), we find
\begin{eqnarray}
C_E({\cal {N}})
&=&\max _{\rho _A\in {\cal {H}}_{in}}
S( \sum _ip_i
|\Phi _A^i\rangle \langle \Phi _A^i|)+
[ S(\sum _ip_i
{\cal {N}}(
|\Phi _A^i\rangle \langle \Phi _A^i|))
-\sum _ip_iS(
{\cal {N}}(
|\Phi _A^i\rangle \langle \Phi _A^i|))]
\nonumber \\
&&-[S(\sum _ip_i\rho _{C'}^i)
-\sum _ip_iS(\rho ^i_{C'})].
\end{eqnarray}
Obviously, the last two terms, the quantity enclosed
in the last $[~]$, give positive value.
One direct corollary of this result is that
the entanglment assisted channel capacity is upper bounded
by the sum of $\log d$ and the one-shot unassisted capacity
proposed by Holevo\cite{H} and proved in Ref.\cite{Fan}
and subsequently proved also in Ref.\cite{H1} by a different method.
Schumacher and Westmoreland \cite{SW2} pointed out that
we can have the following form 
\begin{eqnarray}
C_E({\cal {N}})=
\max _{\rho _A\in {\cal {H}}_{in}}
S(\rho _A)+
C(\{ \rho _{A'}^i\} )
-C(\{ \rho _{C'}^i\} ).
\label{rel}
\end{eqnarray}
where $C(\{ \rho _{A'}^i\} )
=S(\sum _ip_i\rho _{A'}^i)-\sum _ip_iS(\rho ^i_{A'})$,
similarly for $C(\{ \rho _{C'}^i\})$,
and they are related to the Holevo bound\cite{H3}.
Although both $C(\{ \rho _{A'}^i\} )$ and $C(\{ \rho _{C'}^i\})$
depend on the choice of $|\Phi _A^i\rangle $ and the quantum
channel ${\cal {N}}$, the differce 
$C(\{ \rho _{A'}^i\} )
-C(\{ \rho _{C'}^i\} )$ depend only on the overall 
$\rho _A$ as already
being noticed in Ref.\cite{SW2}.
Applications in quantum cryptography 
of the last two terms which is defined as the
coherent information can also be found in Ref.\cite{SW2}. 

In summary, we studied in this paper several aspects of the
entanglement assisted channel capacity such as: data
processing inequalities (\ref{datap}), convexity (\ref{convex}),
additivity (\ref{add}), relationship with one-shot
capacity (\ref{rel}), and property of optimal signal states. 

{\it Acknowlegements:}
The author would like to thank M.Hamada,
K.Matsumoto and other members of ERATO project for
useful discussions, he also would like to thank A.S.Holevo
for useful communications.

\end{document}